# Reuse of Semantic Models for Emerging Smart Grids Applications

Valentina Janev*, Dušan Popadić*, Dea Pujić*, Maria Esther Vidal, Kemele Endris**
* Mihajlo Pupin Institute, University of Belgrade, Serbia
** TIB-Leibniz Information for Centre for Science and
Technology, Hannover, Germany
valentina.janev@pupin.rs, dea.pujic@pupin.rs, dusan.popadic@pupin.rs
maria.vidal@tib.eu, kemele.endris@tib.eu

*Abstract*—Data in the energy domain grows at unprecedented rates. Despite the great potential that IoT platforms and other big data-driven technologies have brought in the energy sector, data exchange and data integration are still not wholly achieved. As a result, fragmented applications are developed against energy data silos, and data exchange is limited to few applications. Therefore, this paper identifies semantic models that can be reused for building interoperable energy management services and applications. The ambition is to innovate the Institute Mihajlo Pupin proprietary SCADA system and to enable integration of PUPIN services/applications in the European Union (EU) Energy Data Space. The selection of reusable models has been done based on a set of scenarios related to electricity balancing services, predictive maintenance services, and services for residential, commercial and industrial sector.

## I. Introduction

Data-driven technologies such as big data and the IoT, in combination with smart infrastructures for management and analytics, are rapidly creating significant opportunities for enhancing industrial productivity and citizen quality of life. As data become increasingly available, the challenge of managing (i.e., selecting, combining, storing, and analyzing) is growing more urgently [1]. Thus, there is a demand for development of computational methods for the ingestion, management, and analysis of big data, as well as for the transformation of these data into knowledge. Semantic intelligence technologies, for instance, are one of the most important ingredients in building knowledge-based systems as they aid machines in integrating and processing resources contextually and intelligently. In order to enable broad data integration, data exchange, and interoperability, and to ensure extraction of information and knowledge, standardization at different levels is needed, e.g., metadata schemata, data representation formats, and licensing conditions of open data. This encompasses all forms of (multilingual) data, including structured and unstructured data, as well as data from a wide range of domains, including geospatial data, statistical data, weather data, public sector information, and research data, to name a few.

The focus in this paper is the energy sector. The energy sector is transforming itself deeply at an extremely fast pace worldwide, moving from the top down vision of energy value chain with centralized production and rigid distribution framework, to the collaborative ecosystem of self-managed prosumers equipped with distributed energy resources and ability to act independently on liberalized energy markets. The recently adopted EU action plans (the European Green Deal [2], the EU Strategy for Energy System Integration [3]) create opportunities for modernization of the energy system, thus making it competitive and sustainable with regard to the environment. In Europe, this paradigm shift has already influenced the entire value chain resulting in: 1) diverse production portfolio with constant increase of renewables penetration; 2) reinforced transmission networks offering higher intra- and cross-border energy exchange and, 3) prosumer-enabled distribution networks with high generation and consumption flexibility. One of the requirements related to procedures for data access in future electricity markets is related to *interoperability of energy services within the Union*. Hence, this paper identifies semantic models that can be reused for building interoperable energy management services and applications. The ambition is to innovate the Institute Mihajlo Pupin (PUPIN) proprietary SCADA system [4] and to facilitate integration of PUPIN services/applications in future integrated energy systems.

The paper is structured as follows. Section 2 presents the motivation, Section 3 discusses the knowledge graph (KGs) creation process and Section 4 presents an example of integration of analytical service with the SCADA system via a semantic layer.

## II. Motivation

### A. Serbian Energy Value Chain

Smart Grids, also referred to as cyber-physical energy systems are the next evolution step of the traditional power grid and are characterized by a bidirectional flow of information and energy. Using the Serbian energy value chain (see an illustration in Figure 1), as an example, we argue that the national electricity infrastructure is not isolated, and hence, interoperability should be ensured in different systems that enable the undisturbed functioning of the electricity production, transmission and consumption. At many parts in the national electricity grid, where the Institute VIEW4 Supervisory control and data acquisition (SCADA) is installed, we have to ensure interoperability with all other applications that exist at client side. The SCADA system is used for monitoring and controlling the energy production, distribution and usage with different objectives including improvement of energy efficiency, increasing the flexibility and renewable generation share, and reducing the energy cost.

B. *Objectives*

The overall goal is to provide an innovative energy management service layer on top of existing supervision and control systems based on reusable semantic models (i.e. a 'knowledge graphs') [5]. The knowledge graph layer will be based on open standards and open APIs and will follow the "no-vendor lock in" principle and ensure that the future services integrate smoothly with different legacy and proprietary solutions. Thus we pose the following research questions:

- RQ1 - Which ontologies cover the needs for modelling the electricity value chain?
- RQ2 - Which ontologies cover the needs for modelling forecasting services?

The goal is to analyze the common schema (vocabularies / ontologies) promoted by the European standards community) and adopt them for the targeted services/applications.

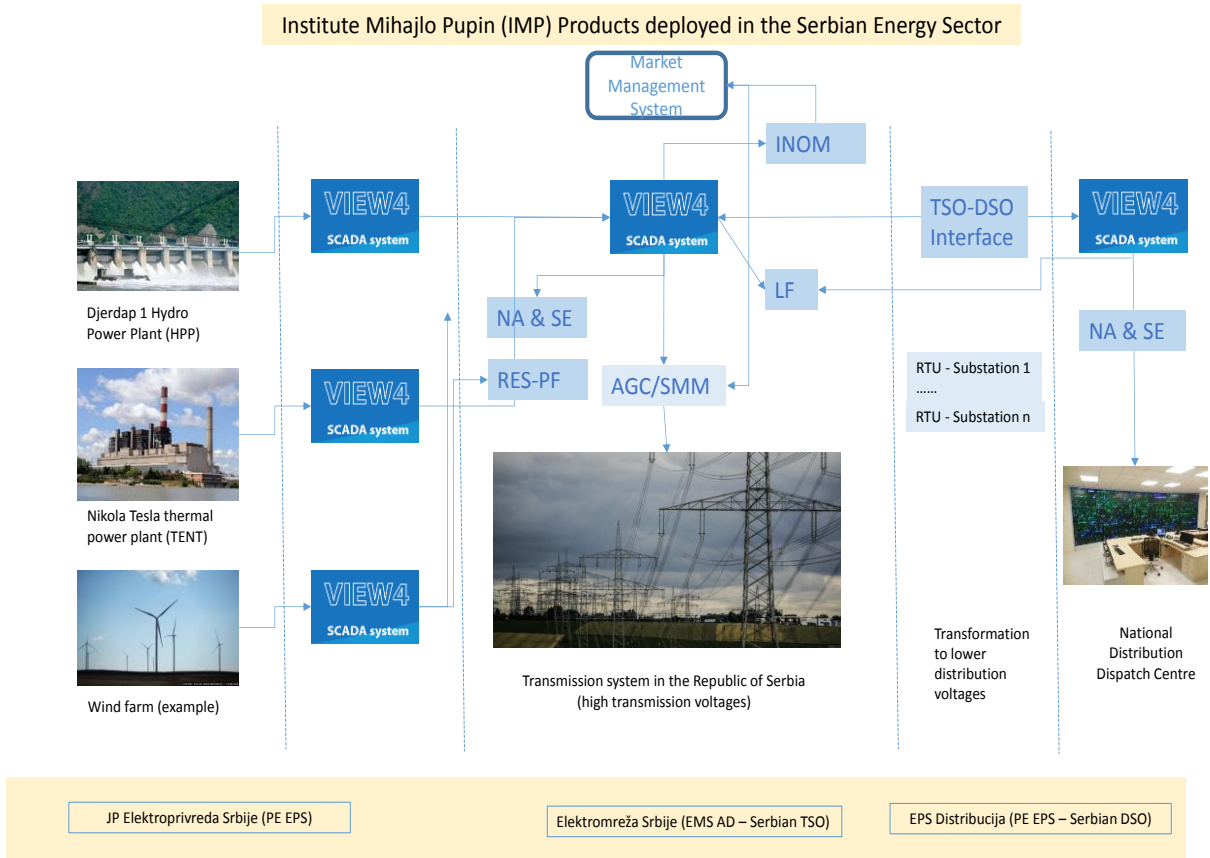

Figure 1. PUPIN SCADA system deployment (examples)

C. *Methodology*

Figure 2 gives an overview of the methodology we follow for building the ARTEMIS interoperability and integration framework [6]. The methodology has extended the approach proposed in the PLATOON project framework for elaboration of semantic layer for Pilot 2a Electricity Balance and Predictive maintenance.

Herein, we will concentrate on the following building blocks from the development of knowledge graph version 1:

- Selection of semantic models;
- Conceptual schema;
- Data connectors;
- Connectors and transformation.

III. KNOWLEDE GRAPH DEVELOPMENT

A. *Selection of semantic models*

Main ontologies considered for reuse in the PUPIN knowledge graph are

- the IEC Common Information Model standards (CIM), see CIM V2.53.0 Schema (MOF, PDF and UML, https://www.dmtf.org/standards/cim/cim_schema_v2530);
- the Smart Appliances REFerence ontology (SAREF), and the extension of SAREF to fully support demand/response use cases in the Energy domain (SAREF4EE);
- the Industrial Data Space (IDS) Information Model, https://international-data-spaces-association.github.io/InformationModel/docs/index.html# ;
- SEAS - Smart Energy Aware Systems, https://w3id.org/seas/ .

## B. Conceptual schema

In the KG conceptualization phase, the knowledge engineer puts different semantic models together in a schema diagram in order to check all classes, properties and possibly to improve them. In this phase, it is important to (1) join the different schema diagram of modules with semantic relations (e.g. subsumption, equivalence, …) and (2) identify the data sources and take actions in order to define the mapping rules. The knowledge engineer also has to evaluate the ontological modules and ensure that its definitions correctly implement the use case requirements and competency questions. The goal of ontology evaluation is to prove compliance of the world model with the world modelled formally. Two important aspects are used for evaluation:

• Competency of the ontology: verify that a representational model is complete with respect to a given set of competency questions.

• Quality requirements: can be measured as the degree of compliance it has with respect to established design criteria (Clarity, Coherence, Modularity).

Examples of competency questions are:
• What are the infrastructures owned by an energy provider?
• Retrieve the current meteorological data for period X
• Retrieve the forecast (production) for period X
• Retrieve the forecast (metrological data) for period X
• Retrieve historical data about production

## C. Data connectors

Data Connector is a component that enables the services to read /write data to production system. In the semantic pipeline, it feeds the Data Preprocessing and Integration components with the raw data. An example of a piece of code from the data connector toward the SCADA system is given below (selecting data from a *photovoltaic solar power plant* database).

| Unique ID column(s) | {plant_id}_{plant_name}_{city} |
|---|---|
| SQL Query | `SELECT plants.id as plant_id,`<br>`       plants.name as plant_name,`<br>`       weather_locations.lat as lat,`<br>`       weather_locations.lon as lon,`<br>`       weather_locations.city as city,`<br>`       assets.asset_name as asset_name,`<br>`       country.country_code as ccode,`<br>`       eic_functions.eic_type_function_acronym as eic_func_acronym,`<br>`       organization.organization_short_name as organization_short_name,`<br>`       organization.organization_name as organization_name`<br>`FROM ``plants```<br>`JOIN weather_locations`<br>`  ON plants.weather_location_id = weather_locations.id`<br>`JOIN assets`<br>`  ON plants.asset_id = assets.id`<br>`JOIN organization`<br>`  ON assets.organization_id = organization.id`<br>`JOIN country`<br>`  ON organization.country_id = country.id`<br>`JOIN eic_functions`<br>`  ON assets.eic_function_id = eic_functions.id`<br>`WHERE  eic_functions.eic_type_function_acronym = 'RES-PV'` |

## D. Semantic transformation and KG creation

Creating a knowledge graph from heterogeneous data sources, for example by merging (1) weather data from Weather API, https://www.weatherbit.io/, and (2) SCADA production data) requires the description of the entities in the data sources using RDF vocabularies, as well as the performance of curation and integration tasks to reduce data quality issues, e.g., missing values or duplicates. Two types of knowledge graph creation strategies: materialized (i.e., data warehousing) and virtual (i.e., Data Lake).

Without going into details how the semantic pipeline was implemented (for more information, please check [7]), herein, we will present examples of reuse of common concepts from the ontologies mentioned before.

**Example 1:** Defining the ARTEMIS-*Plant* ontology

```
artemis:PlantOntology owl:Ontology ;
  dcterms:title "Artemis Plant Ontology"@en ;
  dcterms:issued "2021-05-20"^^xsd:date ;
  dcterms:license
<https://www.apache.org/licenses/LICENSE-2.0> ;
  vann:preferredNamespacePrefix "artemis" ;
  vann:preferredNamespaceUri <https://projekat-artemis.rs/> ;
  owl:versionIRI < https://projekat-artemis.rs/PlantOntology-1.0> ;
  owl:versionInfo "v1.0" .
```

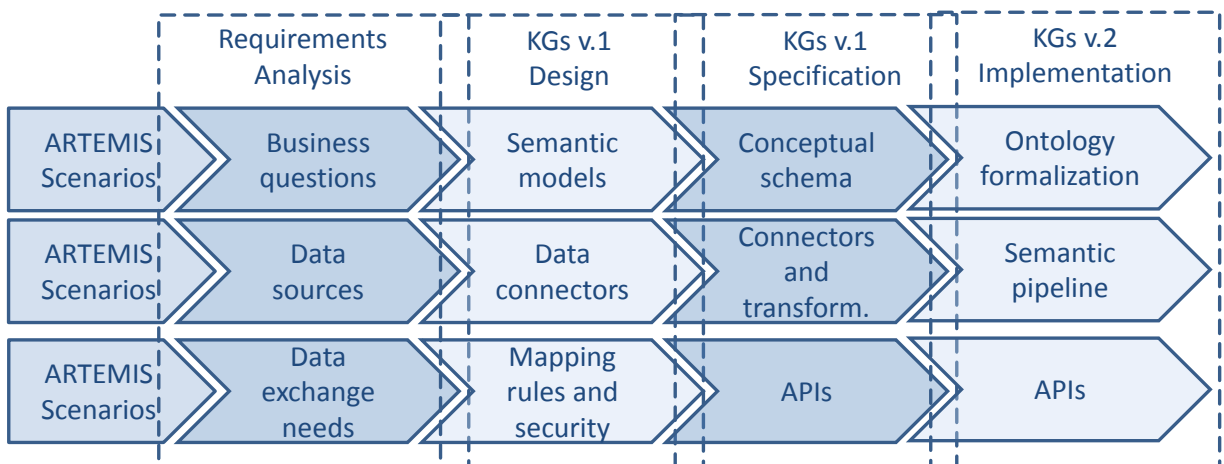

Figure 2. KG development methodology

**Example 2:** Defining a class in the ARTEMIS-*Grid* ontology
```
artemis:ElectricalGrid a owl:Class ;
  rdfs:label "Electrical Grid"@en ;
  rdfs:comment """An electrical grid is an interconnected network for delivering electricity from producers to consumers."""@en ;
  rdfs:subClassOf seas:ElectricPowerSystem;
  vs:term_status "testing" ;
  rdfs:isDefinedBy artemis:GridOntology .
```

**Example 3:** Defining a class in the ARTEMIS-*Energy* ontology
```
artemis:LongTermForecast a owl:Class ;
  rdfs:label "Long Term Forecast"@en ;
  rdfs:comment "The class for long term forecast"@en ;
  rdfs:subClassOf seas:Forecast;
  vs:term_status "testing" ;
  rdfs:isDefinedBy artemis:EnergyOntology .
```

**Example 4:** Defining a dataset
```
artemis:DataSet a owl:Class ;
  rdfs:label "Dataset"@en ;
  rdfs:comment "A data set (or dataset) is a collection of data.  (source: Wikipedia)"@en ;
  rdfs:subClassOf ids:DigitalContent, dcat:Dataset, qb:DataSet;
  vs:term_status "testing" ;
  rdfs:isDefinedBy artemis:Ontology .
```

**Example 5:** Defining a property used in a dataset
```
artemis:hasCapacityActivePower a owl:ObjectProperty ;
  rdfs:label "has capacity active power"@en ;
  rdfs:comment """Links the FeatureOfInterest to its capacity active power property."""@en ;
  owl:subPropertyOf seas:activePower  ;
  rdfs:domain seas:FeatureOfInterest ;
  rdfs:range cim:ActivePower;
   vs:term_status "testing" ;
  rdfs:isDefinedBy artemis:EnergyOntology .
```

The development of a Semantic layer [8] extends the reused common vocabularies and ontologies and the selection of models have to be done based on the target scenarios (e.g. for forecasting, see SEAS ontology).

## IV. SERVICE LAYER DEVELOPMENT

Currently under development are different services that will be tested with the presented semantic layer. Herein, we present an example scenario.

**Scenario 1:** The goal of this scenario is to develop and test a service for more accurate prediction of renewable energy generation (RES forecaster). Different energy supplier strategies for RES should be considered that will support and improve strategic optimization of the resources. The wind power forecasting yields an estimate of the variable power injected in the distribution grid. This allows prediction of when the transformer connecting the distribution grid to the transmission grid will be overloaded, i.e., when local wind turbine generator production will be very high. The various forecasting approaches can be classified according to the type of input (weather prediction, wind turbine generators data, historical production data). Statistically based approaches allow very short-term predictions (2 hours). One of the key challenges for day-ahead forecasting of wind energy remains unscheduled outages that can have large effects on the forecasts for small systems, while the effect is small on the overall grid.

All underlined terms have been modeled in the semantic model with OWL classes and properties. Figure 3 illustrates a use case of using the RES forecaster service with the *Virtuoso* knowledge base.

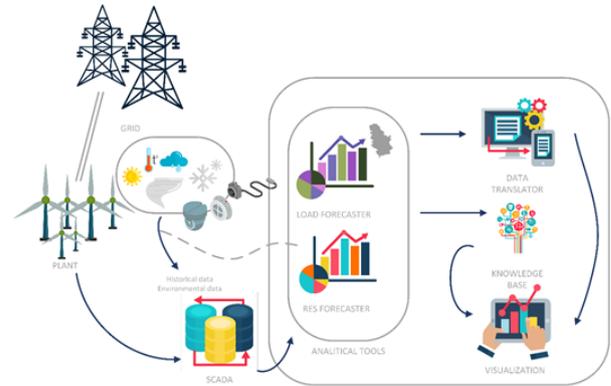

Figure 3. RES service deployment (example)

## V. CONCLUSION

The European electricity system undergoes significant changes driven by the European Union common rules for the internal market for electricity, as well as by the climate action agenda. On a country level, in Serbia, the recently adopted EU energy related strategies create opportunities for modernization of the energy system, thus making it competitive and sustainable with regard to the environment. Additionally, the emerging technologies, edge computing and big data solutions are challenging the traditional energy management applications. Hence, this paper discussed the possibilities of integrating new innovative services that work on top of common semantic models envisioned to be used in future integrated energy value chains.

The selection of an appropriate semantic processing model (i.e., vocabularies, taxonomies, and ontologies that facilitate interoperability) and analytical solution is a challenging problem and depends on the business issues of the targeted domain. In this paper, we showed how to build a semantic model based on common semantic definitions that are found in standards such as CIM, SEAS, SAREF and IDS.

Future work will include development of a second version of the knowledge graph for exposing the PUPIN analytical services in the bigger EU energy data space.

ACKNOWLEDGMENT

This work has been partially supported by the EU H2020 funded projects PLATOON (GA No. 872592), the EU project LAMBDA (GA No. 809965), the EU project SINERGY (GA No. 952140), the EU project TRINITY (GA No. 863874) and partly by the Ministry